\documentstyle[11pt,newpasp,twoside,epsf]{article}
\begin{document}
\title{On the Masses of the Secondary Stars in CVs}
\author{Steve B. Howell}
\affil{Astrophysics Group, Planetary Science Institute, 620 N. 6th Ave., Tucson AZ
85705}

\begin{abstract}
Recent theoretical efforts have made predictions for the masses of the secondary stars
in cataclysmic variables. Accurate observational determinations for M$_2$ are sparse
and typically have uncertainties near $\pm$0.1M$_{\odot}$. How well do theory and
observation agree?
\end{abstract}

\section{Secondary Star Masses}

Theoretical work by Beuermann et al. (1998), Baraffe, and Kolb (2000), and Howell et
al. (2001) and references therein have developed detailed models which
provide mass estimates for CV secondaries as a function of orbital period.
All of these works are in fair agreement and propose that most secondary stars
in systems with short orbital periods ($<$ 5.5 hr) are more or less
the most ``normal" secondary
stars in any CV. Additionally, these authors conclude that as a CV evolves towards
the top of the period gap, the secondary will become bloated, as it is not in thermal
equilibrium, and attain a larger radius for its mass compared with a main sequence
counterpart. Details of the expected observational properties of these bloated 
secondaries can be found in Howell (2001).

Observational work by a number of authors has been compiled by Smith and Dhillon 
(1998) and from that work we find that there are only 16 CVs with orbital periods
under 5.5 hr which have
reliable mass estimates (i.e., 
with well determined uncertainties) for the secondary star. 

Using the theoretical work of Howell et al. (2001) as an example 
for the evolution of CVs, in
particular the evolution and properties of the secondary stars, we plot their
expected relation for M$_2$ as a function of orbital period (dashed line) 
in Figure 1. Also included
(solid line) is the relation expected for main sequence stars of radius equal to the
Roche Lobe radius at a given orbital period. Over-plotted in Figure 1 are the sixteen
reliable M$_2$ masses from Smith and Dhillon.

We can easily see from Figure 1 that at present, the observational data 
can not in general confirm or deny
whether the theoretical models are correct. Only five of the masses have sufficiently
small uncertainties to be useful, and of those, the two above the period gap seem to
favor the theoretical model while below the gap the three well determined values 
fall below
both relations. What is needed to help solve this issue are more, accurate
determinations of the masses of the secondary stars in CVs. These are likely to come
about through the use of high S/N IR spectroscopy obtained with 
large ($>$4m) telescopes.

\acknowledgements
SBH acknowledges partial support for this research from NSF grant
AST 98-19770 and NASA Theory grant NAG5-8500.

%
\begin{figure}
\plotfiddle{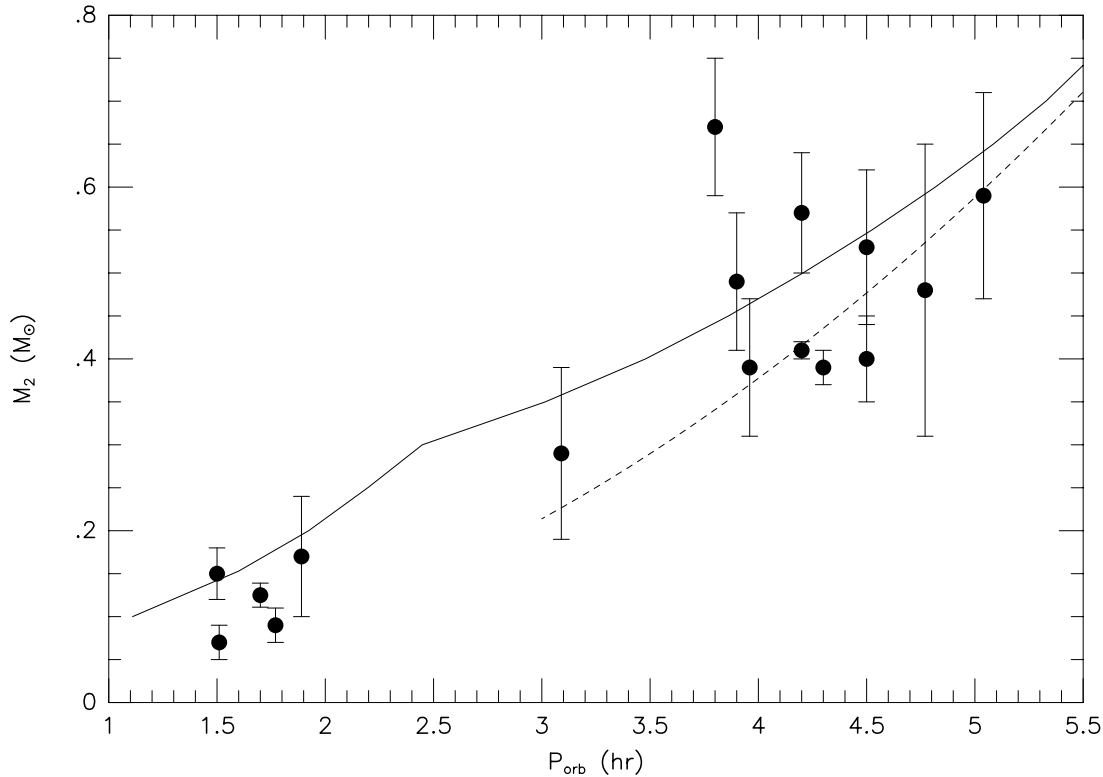}{4.00truein}{90}{60}{60}{235}{-30}
\caption[]{Theory vs. Observation. Plotted are the secondary star mass vs. orbital
period for three sets of data; {\it Solid line} - the mass--orbital period relation for
normal (single) main sequence stars assuming a radius for the star equal to the Roche
Lobe radius of the secondary, {\it Dashed line} - the mass--orbital period 
relation from
Howell et al. (2001) for mass losing CV secondaries, {\it Points} - the 
sixteen reliable
secondary star masses from Smith and Dhillon (1998). We note that the dashed line
essentially matches the solid curve for orbital periods below 2.5 hr.}
\end{figure}

\end{document}